\documentclass[sigconf]{acmart}

\AtBeginDocument{%
  }

\copyrightyear{2025}
\acmYear{2025}
\setcopyright{rightsretained}
\acmConference[SIGCSE TS 2025]{Proceedings of the 56th ACM Technical Symposium on Computer Science Education V. 1}{February 26-March 1, 2025}{Pittsburgh, PA, USA}
\acmBooktitle{Proceedings of the 56th ACM Technical Symposium on Computer Science Education V. 1 (SIGCSE TS 2025), February 26-March 1, 2025, Pittsburgh, PA, USA}
\acmDOI{10.1145/3641554.3701918}
\acmISBN{979-8-4007-0531-1/25/02}

\usepackage{adjustbox}
\usepackage{graphicx}
\usepackage{xcolor}
\usepackage{balance}

\newcommand{\uci}{UC Irvine }
\newcommand{\usu}{Utah State University }

\begin{document}

\title{Construction and Preliminary Validation of a Dynamic Programming Concept Inventory}

\author{Matthew Ferland}
\authornotemark[1]
\orcid{0000-0001-5289-7567}
\affiliation{%
  \institution{Dickinson College}
  \city{Carlisle}
  \state{PA}
  \country{USA}
}
\email{ferlandm@dickinson.edu}

\author{Varun Nagaraj Rao}
\authornote{Both authors contributed equally to this research.}
\orcid{0000-0002-4692-2196}
\affiliation{%
  \institution{Princeton University}
  \city{Princeton}
  \state{NJ}
  \country{USA}
}
\email{varunrao@princeton.edu}

\author{Arushi Arora}
\orcid{0000-0002-9683-8283} %
\affiliation{%
  \institution{University of California, Irvine}
  \city{Irvine}
  \state{CA}
  \country{USA}}
\email{arushia2@uci.edu}

\author{Drew van der Poel}
\orcid{1234-5678-9012}  %
\affiliation{%
  \institution{Northeastern University}
  \city{Boston}
  \country{USA}
}
\email{a.vanderpoel@northeastern.edu}

\author{Michael Luu}
\orcid{1234-5678-9012} %
\affiliation{%
  \institution{University of California, Irvine}
  \city{Irvine}
  \state{CA}
  \country{USA}}
\email{luum6@uci.edu}

\author{Randy Huynh}
\orcid{0000-0002-8506-1245} %
\affiliation{%
  \institution{University of California, Irvine}
  \city{Irvine}
  \state{CA}
  \country{USA}}
\email{randylh@uci.edu}

\author{Freddy Reiber}
\orcid{0000-0003-0007-5072} %
\affiliation{%
  \institution{Boston University}
  \city{Boston}
  \state{MA}
  \country{USA}}
\email{freddyr@bu.edu}

\author{Sandra Ossman}
\orcid{0009-0004-8183-728X}
\affiliation{%
  \institution{University of California, Irvine}
  \city{Irvine}
  \state{CA}
  \country{USA}}
\email{sossman@uci.edu}

\author{Seth Poulsen}
\orcid{0000-0001-6284-9972}
\affiliation{%
  \institution{Utah State University}
  \city{Logan}
  \state{UT}
  \country{USA}}
\email{seth.poulsen@usu.edu}

\author{Michael Shindler}
\orcid{0000-0002-3365-1729} %
\affiliation{%
  \institution{University of California, Irvine}
  \city{Irvine}
  \state{CA}
  \country{USA}}
\email{mikes@uci.edu}

\renewcommand{\shortauthors}{Matthew Ferland \textit{et al.}}

\begin{abstract}

Concept inventories are standardized assessments that evaluate student understanding of key concepts within academic disciplines. While prevalent across STEM fields, their development lags for advanced computer science topics like dynamic programming (DP)---an algorithmic technique that poses significant conceptual challenges for undergraduates. To fill this gap, we developed and validated a Dynamic Programming Concept Inventory (DPCI). We detail the iterative process used to formulate multiple-choice questions targeting known student misconceptions about DP concepts identified through prior research studies. We discuss key decisions, tradeoffs, and challenges faced in crafting probing questions to subtly reveal these conceptual misunderstandings. We conducted a preliminary psychometric validation by administering the DPCI to 172 undergraduate CS students finding our questions to be of appropriate difficulty and effectively discriminating between differing levels of student understanding. Taken together, our validated DPCI will enable instructors to accurately assess student mastery of DP. Moreover, our approach for devising a concept inventory for an advanced theoretical computer science concept can guide future efforts to create assessments for other under-evaluated areas currently lacking coverage.

\end{abstract}

\ccsdesc[500]{Social and professional topics~Computer science education}
\ccsdesc[500]{Social and professional topics~Student assessment}
\ccsdesc[500]{Social and professional topics~Model curricula}

\keywords{
dynamic programming; algorithms education; concept inventory
}

\maketitle

\section{Introduction}

Concept inventories are standardized assessment instruments used in education research to evaluate student understanding of key concepts within a discipline. They are typically comprised of multiple-choice questions targeting common misconceptions identified. These misconceptions are derived via education research. Concept inventories have been developed and validated across many science, technology, engineering, and math (STEM) topics. One of the first and most widely adopted was the Force Concept Inventory (FCI) in physics education \citep{hake1998, fci}. The FCI was devised based on studies of prevalent misconceptions physics students held regarding Newtonian mechanical concepts. It has become a standard instrument for measuring the effectiveness of high school and university physics instruction on improving the conceptual understanding of Newtonian mechanics. Following the FCI's lead, additional concept inventories have been created across other natural science disciplines as well as engineering, including, chemistry \citep{mulford2002inventory}, astronomy \citep{bardar2007development}, biology\citep{anderson2002development}, mathematics \citep{epstein2007development}, statistics\citep{stone2003statistics}, and geosciences \citep{libarkin2005assessment}, among others \citep{murtaza2023taking}.

Within computer science, concept inventories have been formulated for introductory CS1/CS2 courses \citep{danielsiek2012detecting, goldman2010setting}, architecture \citep{porter2013evaluating}, operating systems \citep{webb2014developing}, and cybersecurity \citep{poulsen2021psychometric}. However, concept inventory development remains lacking for more theoretical and advanced CS topics. An unvalidated concept inventory focused on algorithms and data structures was proposed by \citet{paul2013hunting}, but no rigorously validated instrument currently exists for the challenging concept of dynamic programming (DP).

DP is an algorithm design technique, taught in many undergraduate algorithms courses, that builds solutions to problems using optimal substructure and overlapping subproblems. As an advanced algorithmic approach, DP is considered among the most conceptually difficult topics for CS students~\cite{zehra2018student, shindler2022student, enstrom2017iteratively}. Prior work by \citet{shindler2022student} examined prevalent student misconceptions related to dynamic programming, suggesting the need for a concept inventory to assess student understanding. However, despite its reputation as a challenging topic fraught with subtleties students struggle to grasp, and the cataloged misunderstandings they harbor, no concept inventory has yet been formulated and validated to evaluate dynamic programming comprehension.

This paper discusses the development and initial validation of a research-based Dynamic Programming Concept Inventory (DPCI). The DPCI can be deployed by instructors to measure effectiveness of teaching DP to undergraduate computer science majors. We detail the iterative process of crafting valid DPCI multiple choice questions which reveal the presence of student misconceptions identified by prior education studies. Carefully formulating MCQs to subtly tease out conceptual misunderstandings around the intricacies of dynamic programming is difficult and time intensive. We discuss key steps, decisions, tradeoffs and consequences, challenges encountered in producing a high quality concept inventory and preliminary psychometric validation administering our DPCI to 172 undergraduate computer science students.

Our validated DPCI will enable instructors to evaluate student mastery of dynamic programming within undergraduate algorithms courses. Moreover, our experiences devising a concept inventory to measure understanding of an advanced conceptual CS topic will inform development of additional assessments for theoretical areas of computer science presently lacking coverage. With a broad foundation of validated concept inventories across the CS curriculum, instructors can better evaluate the efficacy of pedagogical techniques and curricular interventions. Our contributions include: (i) steps involved in constructing a validated DPCI, (ii) publishing a list of questions to be administered as part of the DPCI,\footnote{DPCI link: \url{https://us.prairielearn.com/pl/public/course/5595/questions}} and (iii) psychometric validation results using classical test theory. Our study was approved as exempt by the IRB.

\section{Related Work}
\subsection{Concept Inventories}
\label{sec:CIs}
A {\it concept inventory (CI)} is a validated assessment for a
topic or set of topics that allows
researchers and instructors to measure what students have learned.
Concept inventory research started with the Force Concept inventory, which allowed for measuring effectiveness of pedagogical techniques against one another in introductory physics classes, helping build the case for more active learning in these courses
\citep{fci,hake1998}. %

Recently, computing education researchers have been creating
CIs, so our discipline can also benefit from them.
Despite their relatively recent introduction in computer science, concept inventories have already been useful for many purposes in computing education research. Introductory programming concept inventories have been used to examine the
relationship between spatial ability and learning programming~\citep{bockmon2020cs1}, to evaluate the effectiveness of teaching students using both block- and text-based
programming languages~\citep{blanchard2020dual}, to understand
the impact of students' educational background on learning topics in computer
science~\citep{ben2017effect}, and to evaluate novel instructional
practices~\citep{nelson2017comprehension,xie2018explicit}.
Concept inventories for other topics in CS have been used to
compare outcomes between digital logic courses which use differing
pedagogical approaches~\citep{herman2010preliminary}, to measure misconceptions of students learning algorithm analysis~\citep{farghally2017towards}, and to evaluate the effectiveness of ways of teaching about the memory model of the Rust programming language~\citep{crichton2023grounded}.
This progress in CS education afforded by the creation and existence of concept inventories gives
a strong argument for their continued development.

Many concept inventories have been created and fully validated for many introductory programming topics, with validated CIs for CS1, first year computer science, middle-grades computer science, object oriented programming, and basic data structures as well as topic-specific CIs for recursion and Java arrays~\citep{murtaza2023taking}.

Theoretical and advanced topics in computer science have not yet received as much coverage in concept inventories. There are fully validated concept inventories only for digital logic and cybersecurity~\citep{herman2014psychometric,herman2023psychometric,poulsen2021psychometric}. Concept inventory work has begun, but not been completed for operating systems, algorithm analysis, and compiler construction~\citep{murtaza2023taking}. This leaves huge gaps in many theoretical and upper-level computing topics regarded as core topics in the ACM curricular guidelines, including all types of algorithm design as well as human-computer interaction, artificial intelligence, networking, and parallel and distributed computing~\citep{acm2013curriculum}. Thus, we find it important that concept inventories are developed for more advanced and theoretical computing topics, to drive the same educational innovation and research that has happened for introductory programming. For extensive reviews of assessment instruments used in
computing education research, see~\citep{decker2019topical,margulieux_review_2019,murtaza2023taking}.

\subsection{Work on Algorithms Education}
While many students begin learning algorithms as a part of their introductory course sequence, the ACM curricular guidelines~\citep{acm2013curriculum} recommend that they take a more advanced algorithms-focused course later on in their curriculum. Such courses almost always cover dynamic programming~\citep{algsCourses}. Dynamic programming is the area of algorithms education with the most research about it~\citep{liu2024teaching}.

Though there is a draft concept inventory for algorithm analysis~\citep{farghally2017towards}, there have not been any concept inventories created to cover the broad space of algorithm design techniques, including but not limited to greedy algorithms, divide-and-conquer algorithms, algorithmic reductions, and dynamic programming (DP) algorithms. Many of these areas of algorithms design are so understudied that creating a concept inventory would not yet be possible until researchers first complete the foundational work of discovering common students misconceptions and struggles in those areas~\citep{liu2024teaching}. On the other hand, studies have been done to understand the nature of student difficulties in dynamic programming~\citep{zehra2018student,shindler2022student,liu2025student}, making it a good topic to design a concept inventory around at this point in time.
Specifically, prior studies have highlighted that students often fail to recognize that DP is the best solution to a problem, and once they do attempt a DP solution, they often fail to select the correct base case or determine the proper overlapping subproblems.

\section{Methods}
\label{sec:methods}

When developing a Concept Inventory, researchers typically follow a predefined set of steps established in previous research on the topic. For our study, we followed the methodology outlined by  \cite{TCWZLP20, adams2011development}. In general, concept inventory studies typically follow a five-step pattern:

\begin{enumerate}
    \item Identify the topics to be addressed in the concept inventory.
    \item Establish common misconceptions held by students in the specific course or subject area.
    \item Create questions designed to identify these misconceptions.
    \item Conduct a validation process in which instructors assess whether the questions effectively identify the presence of misconceptions in the test-takers.
    \item Revise the inventory questions and repeat step 4 as necessary.
\end{enumerate}

Creating an effective concept inventory is a meticulous and time-consuming process that demands a high level of attention to detail. Therefore, our concept inventory must concentrate on fundamental topics, rather than attempting to encompass the entirety of the overarching subject. Given that the assessment is designed to be completed within roughly thirty minutes, it is important to address only a select subset of topics. This ensures that test-takers have enough time to complete the assessment.

    Most concept inventories encompass a wide range of topics. In our case, we chose to focus on dynamic programming and thus completed step one of the process. Furthermore, step two of the process, establishing common misconceptions about dynamic programming has been completed by \citep{zehra2018student} and replicated by \citep{shindler2022student}. Additionally, we have completed step three by creating questions to identify the misconceptions highlighted by our prior work.

\section{Construction of the DPCI}

We attempted to write problems based on  \citet{shindler2022student}, but found it difficult. The problems identified in the paper were primarily about what students did that was incorrect, rather than the student beliefs that caused them to make these mistakes. With IRB approval, we were able to obtain a copy of the 64 anonymized transcripts used in that study, and relabel them to be focused on what student misconceptions were present. We then created an initial set of questions. Then, after testing with students, several questions were removed from the set.

\subsection{Reanalysis of Interview Data}

\begin{table}
    \centering
\begin{tabular}{p{1cm}|p{6.8cm}}
 Number & Description  \\
 \hline
 1 & 2D DP solutions always traverse top left to bottom right  \\
 2 & DP always involves minimization or maximization  \\
 3 & If a problem has minimization or maximization, it must be possible to efficiently solve it with DP  \\
 4 & If a problem has minimization or maximization, there is no DP solution for it  \\
 5 & Conflating recursion with DP  \\
 6 & Memoization is not a DP approach \\
 7 & Aspects of greedy problems are mistakenly associated with DP  \\
 8 & Ignoring specific subproblems (and knowing which ones to solve)  \\
 9 & Polynomial number of subproblems indicates exponential runtime  \\
 10 & DP involves ``non-adjacency'' of subproblems  \\
 11 & Every problem has an efficient DP solution  \\
 12 & DP must memoize every permutation of decisions made in a problem \\
 13 & DP involves tracking aspects of state to remember how a solution was constructed  \\
 14 & DP always requires a 2D array  \\
 15 & DP requires nested for loops, with the inner loop solving a recurrence \\ \hline
\end{tabular}
\caption{List of misconceptions found during re-analysis of interview data, explanations of what they mean, and (maybe) examples.}
\label{tab:misconceptions}
\end{table}

We examined the interview transcripts to search for misconceptions.  So long as a misconception was seen at least once, it was labeled as a misconception, and the numbers found were not quantified. The idea was that uncommon misconceptions would prove to create easy questions that could later be removed.

Anecdotally, the most common misconceptions at this stage were misconceptions 2, 5, and 14.

\subsection{Construction of Questions to Match Misconceptions}
As previously mentioned, initial work on the questions happened before the relabeling that resulted from re-analysis of the interview data. As such, some questions were left over from this initial stage, and were later relabeled to match the updated misconception numbers.

The rest of the questions, however, were generated after the misconception list was identified. First, each author was given full freedom to create as many questions as they wished, targeting at least one misconception identified. Simultaneously, the old questions were relabeled, as mentioned previously, and then removed if they had no match. Then, we identified which misconceptions lacked any questions, and made a question for each of them.

After testing, some questions were rewritten, seperated into multiple parts, or removed, leaving the current questions remaining.

\begin{table}[ht]
\centering
\begin{tabular}{p{1.1cm}||p{1.8cm}|p{2cm}|p{2.2cm}}
 Question ID & Misconception ID & Answer Choice for misconception & \% who Chose UCI/USU \\
     \hline
VR5 & 1 & A & 12 / 14 \\
DV14.1 & 2 & B & --- / 46 \\ %
AA2 & 3 &  D & 14/ 9\\
VR2 & 4 &  A & 5 / 7\\
AA2 & 5 & B & 7 / 7\\
MF3 & 5 &  A, B, or D &  16 / 39 \\ %

DV14.3 & 7 & True & --- / 20 \\

DV6 & 8 & Not A & 13 / 37\\

MF6 & 11 & A &37 / 31\\
AA2 & 11 & A & 3 / 4 \\  %
ML2 & 13 &  A &7 / 30\\ %
DV10 & 14 & A or B& 10 / 5\\
VR2 & 14 & B& 9 / 3\\
DV10 & 15 & A or C& 8 / 7\\

\hline

\end{tabular}
\caption{Frequency at which misconceptions were chosen.  The four questions labeled `other' are questions left over from the initial stage that don't map to one of our core misconceptions, but we found useful to gauge student difficulty with dynamic programming.  Some questions appear in multiple rows;  this indicates that were were different incorrect answer choices that mapped to different misconceptions.}
\label{tab:questions}
\end{table}

\subsection{What we can and cannot test with Multiple Choice}

While multiple choice questions are well-suited for probing conceptual understanding, there are limitations in what they can effectively assess when it comes to dynamic programming. Our DPCI focuses on evaluating mastery of the core principles and requisite reasoning about optimal substructure and overlapping subproblems. However, we decided against including questions that attempt to test other skills like constructing recurrence relations or transition functions. Such procedural tasks are better evaluated through other formats like write-out problems or coding exercises.

\subsection{Pseudocode}
Nine of the questions on the DPCI ask students to answer questions about snippets of code. For our code snippets, we used a pseudocode which was designed to be understandable by students familiar with any common imperative programming language, similar to the approach taken by the SCS1 and BDSI~\cite{porter2019bdsi,parker2016replication}. 
We included a pseudocode guide defining the language constructs at the beginning of the assessment.\footnote{Links to the pseudocode guide and DPCI questions are omitted for anonymouse review but will be included in the final version.}
Our
pseudocode differed from the SCS1 pseudocode in a few key ways. Because we had no need for object oriented constructs, we did not include these in the language. On the other hand, we found that for defining algorithms in a clear manner, it was useful to include various built-in functions such as \texttt{min}, \texttt{max}, \texttt{len}, and \texttt{floor}, which have roughly the same notation as their counterparts in Python. We defined these functions and their asymptotic performance behavior as a part of the pseudocode guide.

\subsection{Outside Expert Review}

We had a draft of our questions by the end of January.  We prepared a draft of the questions on PrairieLearn, which would offer both the question and prompt for feedback on each.  We asked several experts %
to review the questions and provide commentary.  Based on their feedback, we made many changes described below.

Most revisions primarily focused on correcting syntax errors to adhere to our pseudocode guide, refining time complexities, clarifying ambiguities within questions, and ensuring accurate variable assignments. 
Feedback from experts also helped us to ensure the questions focused on testing conceptual knowledge rather than just debugging code.
We also reworded some questions so that students could answer them correctly whether or not their instructor had taught them that recursion with memoization is a valid form of Dynamic Programming (along with this, we are no longer attempting to measure Misconception 6, since not all instructors agree that it is a misconception). 

We also had feedback that the concept inventory didn't cover certain classes of problems, such as dynamic programming on graph structures, or problems where the subproblems are not defined as a suffix or prefix. We felt that it was appropriate to omit these advanced problem types in the interest of focusing on assessing basic dynamic programming concepts. Future work could create a concept inventory for more advanced topics. Finally, we also had feedback that the DPCI did not have questions asking students to create a recurrence for a DP problem, or perform some other technical skill related to DP. We feel that this is appropriate, since as a concept inventory the DPCI is designed to test conceptual knowledge, not necessarily technical skill. Creating assessment questions to test these skills is another avenue for future work in DP assessment.

\section{Preliminary Psychometric Validation}
\subsection{Methods}

Two primary psychometric validation methods, Classical Test Theory (CTT) and Item Response Theory (IRT), have been used in prior work on concept inventories ~\cite{offenberger2019initial,jorion2015analytic,porter2019bdsi}. CTT is utilized when sample sizes are smaller and rapid iteration is necessary. In contrast, IRT is applied under conditions of larger datasets and far more reliable, as it offers robust estimations less influenced by sample specifics. We opted for CTT due to its practicality with our limited sample size and as an initial testing of an assessment instrument.

Reliability within CTT is established using the statistical measure of Cronbach’s $\alpha$, which reflects the internal consistency of the assessment.  A Cronbach's $\alpha$ close to 1 indicates high reliability, with values above 0.8 considered good and above 0.7 satisfactory~\cite{jorion2015analytic}.

To assess question validity, we also calculate difficulty and discrimination. Difficulty is defined as the proportion of examinees who answer a question correctly, with an ideal range from 0.2 to 0.8, ensuring a balanced challenge. Discrimination is measured by the point-biserial correlation between a questions score and the total test score~\cite{jorion2015analytic}, with values above 0.2 indicating effective discrimination between differing levels of student understanding.

\subsection{Data Collection}
Students in the algorithms course at \uci{} were provided the DPCI as an optional practice exam.  The course covered dynamic programming in class for two weeks, and the DPCI optional practice exam was made available three days before an in-class quiz on the topic.   Of the $316$ students in the class, $109$ completed the exam, and of those, $93$ agreed to allow us to use their data for research.   That quarter, students were expected to complete 19 warm-up activities before various lecture meetings as part of their grade;  students who took the DPCI practice exam, regardless of whether or not their data was used for research, were allowed to drop one warm-up activity from their course grade.

A few weeks later after some minor revisions to the DPCI questions (detailed in Section~\ref{sec:revisions}), students at \usu{} were provide the DPCI as homework assignment to complete (graded only on completion) before their midterm on Dynamic Programming. Out of the $73$ students in the class, $63$ completed the exam, and the terms of our IRB allowed us to use all of their data.

\subsection{Results}
\subsubsection{Round 1}
Our analysis of question difficulties revealed a range from 0.20 to 0.84, with the exception of item Q3, which presented a notably lower difficulty of 0.09 (see Table \ref{tab:difdis}). A lower difficulty score indicates a more challenging question, and while most items fell within the accepted difficulty range of 0.20 to 0.80, DV14 was significantly more difficult. This variation in difficulty levels suggests that the DPCI offers a balanced mix of easy, medium, and hard questions, catering to students with varying proficiency in dynamic programming concepts.

The discrimination values ranged from 0.28 to 0.56, indicating that most questions effectively differentiate between students of differing skill levels (see Table \ref{tab:difdis}). However, questions DV13 and DV14 were outliers, with discrimination values of -0.08 and 0.13, respectively, falling outside the recommended minimum of 0.20. Except for these two questions, the discrimination values provide substantial evidence that higher question scores correlate positively with overall proficiency as measured by total performance on the assessment.

We obtained a Cronbach's $\alpha =0.76$. The Cronbach's $\alpha$ after removing questions DV13 and DV14 was $0.78$. This value is close to \citet{jorion2015analytic} recommendation ($\ge 0.8$) for good reliability and is strong
when compared to published values of other CIs.

\subsubsection{Revisions based on Round 1}
\label{sec:revisions}

We made several modifications to questions in the CI based on our initial round of validation. Our primary focus was addressing the two problems with discrimination values below $20\%$.

Problem DV14, the most difficult problem and the second least discriminating, was split into three new problems. Originally a ``select all" style problem with 5 possible choices, we determined it was likely this problem had such a low correctness score simply due to the binary grading scheme. Even if a student correctly selected/didn't select four of the five choices, they were still marked with a zero. To rectify this, we opted to split the problem into three new problems, with each new problem focused on a particular theme (e.g. DP and optimization). The hope was that this would allow us to determine which (if any) of the themes of the problem were causing such a high difficulty metric.

For problem DV13, which was the second most difficult problem and had a negative discrimination value, we simplified the problem by only using asymptotic analysis in the latter part of the implication. We also referred explicitly to the number of elements filled in in the DP table during the course of the algorithm, as opposed to the asking about its asymptotic runtime.

In addition to our edits based on the student metrics, we also made several changes based on student feedback (gathered through comment boxes on PrairieLearn). These included ensuring consistent capitalization of variable names and adding clarifying details.

\subsubsection{Round 2}
The results of the second round of validation can also be seen in Table~\ref{tab:difdis}. The results were largely similar, with some variation that is expected when using the same questions on a different student population from a different university. While the newly created questions DV14.1 and DV14.3 performed well, DV13 and DV14.2 had very poor discrimination measures.

For the second round, we obtained a Cronbach's $\alpha =0.76$. The Cronbach's $\alpha$ after removing questions DV13 and DV14.1 was also $0.76$. This value is similar to that previously obtained,  close to \citet{jorion2015analytic} recommendation for good reliability and is strong when compared to published values of other CIs.

\begin{table}[ht]
\centering
\begin{adjustbox}{max width=\columnwidth}
\begin{tabular}{c|cc|cc||c|cc|cc}
  & \multicolumn{2}{c|}{Round 1} & \multicolumn{2}{c||}{Round 2}
&& \multicolumn{2}{c|}{Round 1} & \multicolumn{2}{c}{Round 2}\\
  \hline
Q & Dif. & Dis. & Dif. & Dis. & Q & Dif. & Dis. & Dif. & Dis.\\
  \hline
MF1 & 0.46 & 0.31 & 0.48 & 0.33 &
  DV4 & 0.80 & 0.48 & 0.81 & 0.43 \\
MF3 & 0.83 & 0.28 & 0.62 & 0.41 &
  DV5 & 0.70 & 0.57 & 0.73 & 0.47 \\
MF6 & 0.62 & 0.29 & 0.68 & 0.35 &
  DV6 & 0.70 & 0.56 & 0.64 & 0.36 \\
AA2 & 0.76 & 0.52 & 0.78 & 0.29 &
  DV7 & 0.31 & 0.42 & 0.29 & 0.32 \\
AA3 & 0.46 & 0.46 & 0.44 & 0.49 &
  DV8 & 0.27 & 0.40 & 0.27 & 0.37 \\
ML2& 0.53 & 0.37 & 0.48 & 0.46 &
  DV9 & 0.24 & 0.31 & 0.40 & 0.39 \\
RH1 & 0.25 & 0.34 & 0.41 & 0.38 &
  DV10 & 0.84 & 0.53 & 0.91 & 0.65 \\
VR2 & 0.74 & 0.43 & 0.86 & 0.49 &
  DV11 & 0.81 & 0.40 & 0.82 & 0.68 \\
VR3.2 & 0.47 & 0.46 & 0.33 & 0.31 &
  DV12 & 0.59 & 0.45 & 0.70 & 0.30 \\
VR2 & 0.40 & 0.38 & 0.43 & 0.16 &
  DV13 & 0.20 & -0.08 & 0.21 & 0.09 \\
VR5 & 0.83 & 0.34 & 0.81 & 0.38 &
  DV14 & 0.09 & 0.13 &---&---\\
DV1 & 0.63 & 0.40 & 0.59 & 0.34 &
  DV14.1 &---&---& 0.13 & 0.40 \\
DV2 & 0.35 & 0.43 & 0.17 & 0.37 &
  DV14.2 &---&---& 0.81 & 0.11 \\
DV3 & 0.75 & 0.46 & 0.82 & 0.51 &
  DV14.3 &---&---& 0.79 & 0.37 \\
    \hline
\end{tabular}
\end{adjustbox}
\caption{Difficulty and Discrimination numbers for data collection at \uci{} (Round 1) and \usu{} (Round 2). Most of our questions fall within the preferred range of discrimination greater than 0.2 and difficulty between 0.2 and 0.8.}
\label{tab:difdis}
\end{table}

\begin{figure}
    \centering
    \includegraphics[width=\columnwidth]{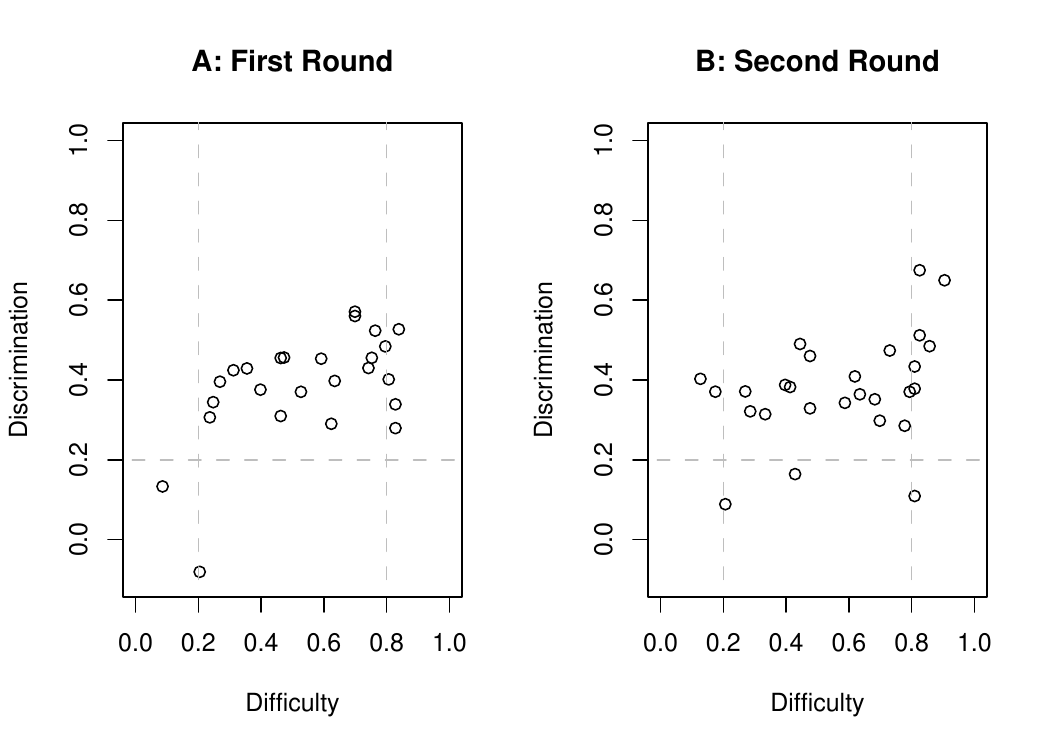}
    \caption{Difficulty and discrimination for both rounds of validation.}
    \Description[A visualization of difficulty and discrimination for both rounds of validation.]{A visualization demonstrating almost all questions have both a difficulty value between 0.2 and 0.8, and a discrimination value of at least 0.2 for both rounds of validation.}
    \label{fig:difdis}
\end{figure}

\section{Discussion}

\subsection{Overall Question Quality}
Overall, our questions performed well with 9 out of the 15 misconceptions being selected by 40\% or more of the students. However, as the results indicated, DV13 and DV14.2 continued to perform poorly even after revisions. Consequently, we removed these questions from the concept inventory, as they proved to be outliers.

\subsection{Evidence that the Misconceptions Are Measured by the Questions}

Misconceptions chosen by 15\% or more of the students provide strong evidence that the misconception both targets and accurately measures the intended misunderstanding, thus validating our hypothesis about its prevalence. 
Highly prevelant misconceptions were:
\begin{itemize}
    \item Misconception 2: DP always involves minimization or maximization
    \item Misconception 5: Conflating recursion with DP 
    \item Misconception 7: Aspects of greedy problems are mistakenly associated
with DP
    \item Misconception 8: Ignoring specific subproblems (and knowing which ones to solve) 
    \item Misconception 11: Every problem has an efficient DP solution 
    \item Misconception 13: DP involves tracking aspects of state to remember how a solution was constructed
\end{itemize}

Low-prevalence misconceptions (less than 15\%) may suggest questions that fail to present the misconception clearly, or that our hypothesized misconception is less widespread than anticipated. 
Low-prevalence misconceptions were:
\begin{itemize}
    \item Misconception 1: 2D DP solutions always traverse top left to bottom right
    \item Misconception 3: If a problem has minimization or maximization, it must be possible to efficiently solve it with DP 
    \item Misconceptions 4: If a problem has minimization or maximization, there is no DP solution for it
    \item  Misconception 14: DP always requires a 2D array 
    \item Misconceptions 15: DP requires nested for loops, with the inner loop solving a recurrence
\end{itemize}

By analyzing selection rates across all categories, we evaluate each question’s construct validity, assessing its ability to measure the intended misconception genuinely. This analysis helps guide refinements to ensure our concept inventory reliably identifies and distinguishes between various misconceptions. 

\section{Conclusions and Future Work}

In this work, we detail our process of developing and validating the first Dynamic Programming Concept Inventory (DPCI). We administered the DPCI to 172 undergraduate computer science students across two large public universities in the U.S., and our psychometric validation revealed that the majority of difficulty and discrimination values across questions fell within acceptable parameters. This indicates that the questions part of the DPCI were of appropriate difficulty and effectively distinguished between students of varied levels of conceptual understanding.

Future research could extend the DPCI's validation across additional universities,
 broadening our understanding of student misconceptions in dynamic programming. This initial step in identifying misconceptions will inform future research on targeted pedagogical interventions. Ultimately, by identifying and evaluating these interventions' impact on learning outcomes, we can refine instructional methodologies to enhance student comprehension of dynamic programming concepts.

\section* {Acknowledgements}

We are extremely grateful to the 93 students at \uci who took a draft of the DPCI and allowed us to use their data for research, and to the 63 students at \usu for the same.  We are also grateful to Jeff Erickson, Dan Hirschberg, David Kempe, and Jonathan Liu, who provided outside expert reviews and commentary. Varun Nagaraj Rao was supported in part by a Graduate Fellowship of Social Data Science by the Data Driven Social Science Initiative at Princeton University.

\bibliographystyle{ACM-Reference-Format}
\balance
\bibliography{references}

\end{document}